\documentclass[oneside,12pt]{article}
\usepackage[utf8]{inputenc}
\usepackage[top=1in,bottom=1in,left=1in,right=1in]{geometry}
\usepackage{authblk}
\usepackage[sort&compress]{natbib} 
\usepackage{graphicx}
\usepackage{placeins}
\usepackage{caption}
\usepackage{subcaption}
\usepackage{float}
\usepackage{amsfonts, amsmath, amssymb, amsthm, constants, bbm}
\usepackage{mathrsfs}
\usepackage{enumitem}
\usepackage{booktabs}
\usepackage{authblk}
\usepackage{microtype}
\usepackage{titlesec}
\usepackage{ragged2e}
\usepackage{array}

\newtheorem{theorem}{Theorem}
\newtheorem{lemma}{Lemma}
\newtheorem{rem}{Remark}

\usepackage{xcolor}
\definecolor{darkred}{RGB}{100,0,0}
\definecolor{darkgreen}{RGB}{0,100,0}
\definecolor{darkblue}{RGB}{0,0,150}

\usepackage{hyperref}
\hypersetup{colorlinks=true, linkcolor=darkblue, citecolor=darkblue, urlcolor=darkblue}
\usepackage{xurl}

\newcommand\numberthis{\addtocounter{equation}{1}\tag{\theequation}}

\newcommand{\emdash}{\unskip\,---\,\unskip}

\input{notation.tex}

\usepackage[T1]{fontenc}
\usepackage{lmodern}
\makeatletter
\newcommand{\twobar}{/\kern-0.2em/} % nice "//"
\let\orig@Url@acthash\Url@acthash % store original \Url@acthash
\let\new@Url@acthash\Url@acthash % make new \Url@acthash that acts "//" as well
\g@addto@macro{\new@Url@acthash}{\Url@Edit\Url@String{//}{\twobar}}
\let\orig@urlstyle\urlstyle % make \urlstyle use the original \Url@acthash
\def\urlstyle{\let\Url@acthash\orig@Url@acthash\orig@urlstyle}
\g@addto@macro{\url@rmstyle}{\let\Url@acthash\new@Url@acthash} % make selected url styles use the new \Url@acthash
\g@addto@macro{\url@sfstyle}{\let\Url@acthash\new@Url@acthash}
\makeatother

\titlespacing{\paragraph}{0pt}{4pt}{4pt}
\titlespacing{\section}{0pt}{8pt}{2pt}
\setlist{nolistsep}

\begin{document}

\title{Efficient Sampling in Disease Surveillance through Subpopulations: Sampling Canaries in the Coal Mine}
\newcommand{\runtitle}{Efficient Sampling in Disease Surveillance}
\author[1]{Ivo V. Stoepker}
\affil[1]{\small Department of Mathematics and Computer Science, Technische Universiteit Eindhoven, Eindhoven, The Netherlands} 
\date{}
\maketitle
\thispagestyle{empty}

\begin{abstract}
\vspace{-0.2cm}
We consider outbreak detection settings of endemic diseases where the population under study consists of various subpopulations available for stratified surveillance. These subpopulations can for example be based on age cohorts, but may also correspond to other subgroups of the population under study such as international travellers. Rather than sampling uniformly across the population, one may elevate the effectiveness of the detection methodology by optimally choosing a sampling subpopulation. We show (under some assumptions) the relative sampling efficiency between two subpopulations is inversely proportional to the ratio of their respective baseline disease risks. This implies one can increase sampling efficiency by sampling from the subpopulation with higher baseline disease risk. Our results require careful treatment of the power curves of exact binomial tests as a function of their sample size, which are non-monotonic due to the underlying discreteness. A case study of COVID-19 cases in the Netherlands illustrates our theoretical findings.

\vspace{0.3cm}
\end{abstract}

\section{Introduction} \label{sec:intro}

Disease outbreak detection is a critical component in public health management \citep{Lombard2007} and as a result a substantial body of research has been dedicated to the development of various outbreak detection methodologies based on public health data (see e.g. \cite{Dato2004, Unkel2012}). However, the effectiveness of such methods hinges crucially on the data sampling strategy, which has received far less attention. This lack of scrutiny leads to inefficient sampling practices and an unwarranted pessimistic attitude towards some common convenience sampling approaches. Motivated by this pessimism, we formally discuss efficient sampling strategies in the context of disease outbreak detection. 

For the practitioner, the headline of our discussion is that if subpopulations are available for stratified surveillance of endemic diseases, then for outbreak detection it can be much more efficient (i.e. requiring fewer samples for equal detection performance) to sample from a subpopulation rather than using a representative sample from the full population. Specifically, subpopulations in which subjects have, under non-outbreak settings, the highest risk of contracting the disease under study may lead (under some assumptions) to considerably more efficient inference if their risk is relatively large in comparison to the average population risk \emdash and we quantify the magnitude of this sampling efficiency in this work. Such high-risk subpopulations may represent the proverbial \emph{``canaries in the coalmine''}, exhibiting more pronounced evidence for disease outbreaks than the full population.

Our discussion spans two surveillance settings of endemic diseases. We start our discussion with a static setting where we sample subjects from a population, uncovering if the subject is currently carrying a disease under study, in order to infer if the disease prevalence in the population exceeds a predetermined nonzero baseline prevalence. We then show how our insights from the static setting extend to a monitoring setting, where a sample of subjects is monitored over time, and data arrives sequentially in the form of daily new cases within this sample. The goal is then to raise an alarm when the average rate of subjects getting diseased in the population exceeds a known baseline rate. 

We consider settings where the population under study consists of multiple subpopulations (which may overlap) indexed by $j=1,\dots,s$, of size $m_j$, available for stratified surveillance, and with a subpopulation-specific baseline disease risk. Outbreaks are assumed to manifest as elevated number of disease cases across all subpopulations and proportional to the baseline disease risk. We revisit and generalize this assumption in the subsequent sections.

\paragraph{Contributions.} For the purposes of outbreak detection, we characterize the sampling efficiency between samples from distinct subpopulations. For the static surveillance setting, a suitable outbreak detection methodology uses the exact binomial test for proportions. One of our contributions and central to our discussion lies in a novel non-asymptotic analytic quantification of the relative sampling efficiency between such exact binomial tests. The ensuing analysis is nontrivial due to the non-monotonic and erratic nature of the binomial test power curve as a function of the sample size \citep{Chernick2012}. We avoid normal approximations, as these have poor accuracy in our regime of interest. The obtained relative sampling efficiency implies (under some assumptions) it is more efficient to sample from subpopulations with a high baseline prevalence. We show how this efficiency result can be extended to a monitoring setting under a common distributional approximation. Finally, an analysis of COVID-19 data further expands on the discussed monitoring setting, corroborating applicability and potential gains of such sampling efficiency in practical settings.

Apart from potentially being more efficient, subpopulations consisting of individuals with a high baseline risk of contracting a disease may be more readily available for sampling as well, as such individuals may be in greater contact with healthcare professionals. This is one reason why our insights are practically relevant and applicable in various contexts. We further discuss two relevant example contexts:

\begin{itemize}[noitemsep]
\item \textit{Travellers to tropical destinations.} In the context of tropical disease surveillance, travellers to tropical destinations can (temporarily) be modelled as a subpopulation of their destination country. Due to increased close contact with other individuals \citep{Freedman2006} or low natural resistance to diseases endemic in their destination \citep{Wilson2003}, for some diseases travellers may have an elevated baseline risk of disease when compared to the domestic population. Therefore travellers are a good example of a subpopulation which may be much more efficient to monitor for the purpose of outbreak detection when compared to the domestic population itself. This contradicts the common pessimistic sentiment that monitoring travellers (as opposed to monitoring the full population) is merely a \emph{convenient} monitoring approach \citep{Fukusumi2016, TaylorSalmon2024, Hamer2020, Leder2017, Bao2022}.

\item \textit{Monitoring of low-risk diseases.} When monitoring diseases with low individual risk of complications (e.g. influenza or COVID-19) many individuals do not communicate infections to health care professionals \citep{Lau2021, Wang2022}, obfuscating potential upward trends in case numbers. Instead of aiming to elevate overall reporting rates, it may be more efficient to focus such efforts on subpopulations for which baseline risk of contracting disease is known to be highest. We discuss a real-world COVID-19 example that underscores this in Section~\ref{sec:case_study}.
\end{itemize}

\paragraph{Related work.} We quantify sampling efficiency of subpopulations in the context of outbreak detection of diseases \emph{endemic} to the population: for example, surveillance of COVID-19 cases to detect increases in the rate at which individuals get diseased. Previous literature \emdash adopting the nomenclature ``risk-based surveillance''\footnote{The word ``risk'' in ``risk-based surveillance'' refers here to the risk of contracting the disease under study. There exists literature adopting similar nomenclature where ``risk'' refers to the \emph{consequences} of disease should they occur in various subpopulations. To disentangle these two concepts some authors have adopted ``risk-based sampling'' when referring to sampling settings and ``risk-based surveillance'' when referring to consequences of outbreaks \citep{Cameron2012}.} or ``targeted surveillance'' \emdash has instead focused on the setting of showing freedom of disease for \emph{emerging} diseases. In the latter setting, \cite{Ferguson2014, Mastin2017} discuss sampling efficiency between vector and host populations. Crucially, for detection of emerging diseases there is no uncertainty upon discovery of a single case: the presence of the emerging disease can then be declared with absolute certainty. In the endemic setting we study (where one aims to detect outbreaks relative to some baseline nonzero disease pressure) discovery of some cases is expected also under non-outbreak scenarios, such that outbreaks are instead declared under uncertainty. To quantify sampling efficiency in such endemic settings, one needs account for this uncertainty as well, and as a result our analysis is of a different nature than that of previous literature. In \cite{Konikoff2015} determining sample sizes in a setting closely connected to our endemic disease setting is discussed, but rather than outbreak detection, the goal there is population prevalence estimation (in the context of cross-sectional HIV studies). Sample sizes requirements are provided for desired (approximate) confidence interval widths, and for power in detecting a change in prevalence in a two-sample setting. The latter results are provided numerically, and are not used to derive insights on sampling efficiency when subpopulations are available for stratified surveillance.

To the best of our knowledge, all earlier works discussing risk-based surveillance pertain the setting of declaring freedom of disease for emerging diseases, with an emphasis on veterinary applications.
A review of risk-based surveillance approaches is provided by \cite{Stark2006}; see also \cite{Doherr2001a, Sergeant2011}. 
The interdisciplinary RISKSUR project \citep{RISKSURConsortium2015} aimed to develop (cost-effective) risk-based surveillance tools.
The technical report \cite{efsa} describes guidelines for risk-based surveillance for surveys of plant pests.
In \cite{Martin2007a} an approach to risk-based surveillance based on scenario trees is discussed.
In \cite{Williams2009a} the impact of estimation uncertainty of disease risk is discussed in the context of risk-based surveillance.
In \cite{Williams2009} a risk-based sampling approach based on Poisson sampling is discussed.
In \cite{Cameron2012, Wells2009} examples are presented which highlight potential efficiency gains of risk-based surveillance. 

\section{Static population sampling}\label{sec:settings-static}
Consider a static setting where an unknown number of subjects within a population is carrying a disease and the goal is to infer, through a sample from the population, if the proportion of diseased subjects is larger than some predefined maximum baseline proportion. Our goal is to identify which subpopulation would be most efficient for this purpose. In the context of this static setting, subpopulations with a higher disease risk have a larger proportion of diseased subjects.

Denote with $x_{i,j}$ the disease status of subject $i$ from subpopulation $j$, with $x_{i,j} = 1$ if subject~$i$ carries the disease and zero otherwise. Denote the fraction of subjects carrying the disease within subpopulation $j$ under a non-outbreak scenario by $p_j$ (i.e. the baseline prevalence) which we assume is known. Denote $m_j$ the size of subpopulation $j$, which we assume is much larger than potential sample sizes of interest. Denote the unknown actual fraction of diseased subjects by $q_j = m_j^{-1}\sum_{i=1}^{m_j} x_{i,j}$. Our aim is now to test, for a subpopulation $j$, the following pair of competing hypotheses:
\begin{equation}\label{hyp:static}
H_0^{(j)}: q_j \leq p_j \ , \quad\text { vs. }\quad H_1^{(j)}: q_j > p_j \ . 
\end{equation}
We make no assumptions on the risk profiles of subjects $x_{i,j}$ at an individual level, but we assume these are unknown and cannot be leveraged for sampling purposes. Therefore for our testing problem in~\eqref{hyp:static} one can do no better than sampling uniformly over the subpopulation. Subsequently, this heterogeneity of risk profiles plays no role in the distribution of the sampled observations, reducing the problem to a binomial testing scenario.\footnote{In rigor, if subjects are sampled without replacement, the samples from the population follow a hypergeometric distribution. Modeling samples through a binomial distribution is a common approximation in such settings (see e.g. \citet{Sandiford1960}), which is accurate when the population size $m_j$ is large compared to the sample size $n_j$ (e.g. in \cite{Brunk1968} it is mentioned the approximation is considered good when $m_j > 10n_j$). Since we assume $m_j$ is much larger than potential samples as well, we align with common practice and do not account for the effects of this approximation.} 

From subpopulation $j$, we consider a sample of size $n_j$ as $(X_{i,j})_{i=1,\dots,n_j}$ where we model $X_{i,j} \sim \text{Bernoulli}(q_j)$. Denote $F_{n,p}$ the cumulative distribution function of the binomial distribution with count parameter $n$ and probability parameter $p$, and $F^{-1}_{n,p}(u)$ its quantile function, i.e.
\begin{equation}\label{eq:quantile_binom}
F_{n,p}^{-1}(u) \equiv \inf\Big\{ x \in \mathbb{R} : F_{n,p}(x) \geq u \Big\} \ .
\end{equation}
Now let $\psi_{j,n,\alpha} : \bbR^n \to \{0,1\}$ denote the exact\footnote{Exactness refers here to the usage of the \emph{exact} null distribution of the observations for calibration, rather than a normal approximation. Due to the discreteness of the observations the resulting test may be conservative; see also Remark~\ref{rem:ump}. The phrasing we use is common in literature \citep{Chernick2012} or software \citep{RCoreTeam2024}, but a more descriptive name could be the non-randomized uniformly most powerful test of level at most $\alpha$.
} binomial test at significance level~$\alpha$ for the hypotheses in~\eqref{hyp:static} using a sample of size $n$ from subpopulation~$j$, i.e.
\begin{equation}\label{eq:exact_bin_test}
\psi_{j,n_j,\alpha}(X_{1,j},\dots,X_{n_j,j}) \equiv \ind{ \sum_{i=1}^{n_j} X_{i,j} > F_{n_j,p_j}^{-1}(1-\alpha)} \ .
\end{equation}
The quantity of interest is now, for each subpopulation, the minimum required sample size such that the power of the binomial test using this sample (under the alternative) is at least some prescribed $1-\beta$ when conducted at significance level $\alpha$. In other words, we investigate the sampling efficiency between these tests when sampling from distinct subpopulations.

\begin{figure}[!htb]
\centering
\includegraphics[width=0.45\textwidth]{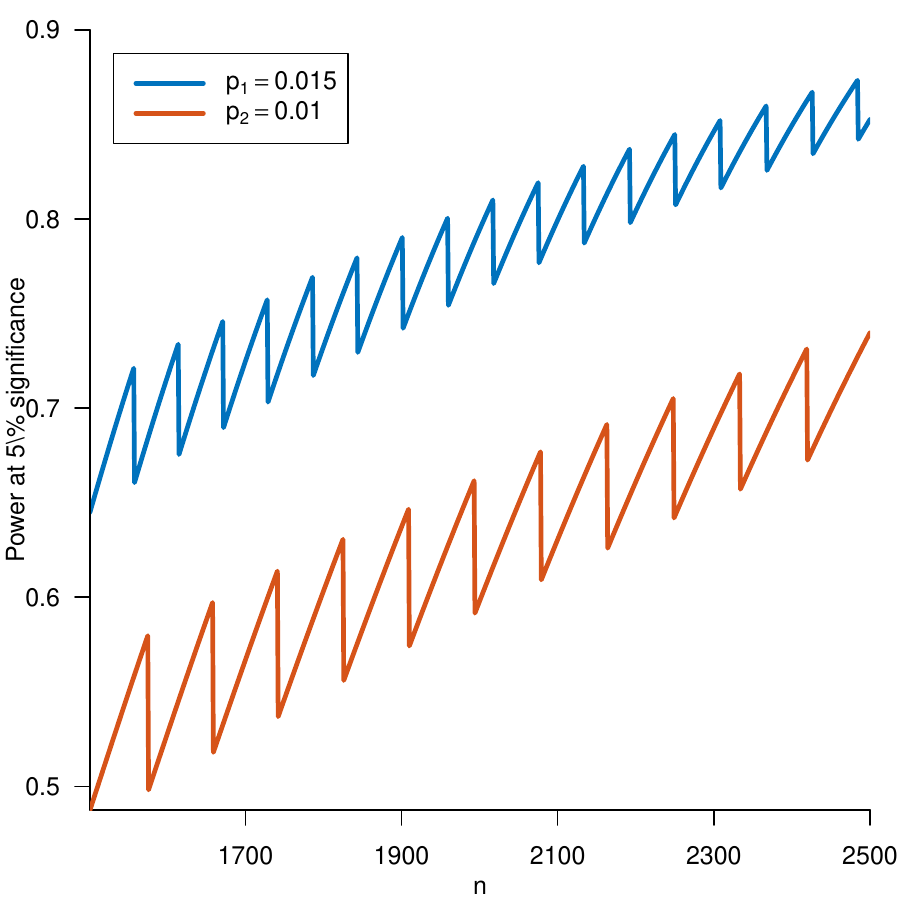}
\caption{Power at 5\% significance of the binomial test~\eqref{eq:exact_bin_test} as a function of the sample size $n$ under the alternative as in~\eqref{hyp:static} where $q_j = 1.5p_j$ for $j=1,2$. Two curves are depicted corresponding to the null $p_1 = 0.015$ and $p_2 = 0.01$ respectively. The curves are only defined for integer $n$, but depicted as smooth lines for legibility.}\label{fig:power}
\end{figure}

At this point, one may be tempted to proceed with arguments under normal approximations for the test~\eqref{eq:exact_bin_test}, for example based on an effect size such as Cohen's $h$ \citep{Cohen1988}. However, the present context implies we are interested in small values of $p_j$ and $q_j$ which render such approximations less accurate, even at sample sizes larger than commonly understood sufficient \citep{Brown2001, Brown2002, Chernick2012}. To enunciate this, Figure~\ref{fig:power} depicts the power of the test~\eqref{eq:exact_bin_test} as a function of the sample size $n$ at a fixed significance level of $\alpha=0.05$. The power is depicted under two alternatives with different associated null hypotheses: $p_1 = 0.015$ and $q_1 = 1.5p_1$, and $p_2 = 0.01$ and $q_2 = 1.5p_2$. The shape of the power curves is non-monotonic, resembling a ``saw-tooth'' \citep{Chernick2012}. 

The results in Figure~\ref{fig:power} highlight two things. First, they enunciate that the associated normal approximations may not be sufficiently accurate in this setting: observe that the size of the ``teeth'' in Figure~\ref{fig:power} can be larger than $8\%$. Naturally, it depends on the context whether such inaccuracies are problematic. Nevertheless, the figure depicts the power of tests at a sample size of $n > 1500$ and $np_1 > 15$ \emdash these are arguably not small-sample curiosities. Secondly, the results underscore that relating these two power functions is not straightforward: in particular, one must be careful to account for misalignment of the ``teeth''. In fact, from this figure it is not clear if a meaningful relationship between these two power functions even exists.

In this paper we show that the power curves in Figure~\ref{fig:power} can be meaningfully related. 
Our technical arguments must carefully account for the underlying non-monotonic nature of the exact binomial test power curve as a function of the sample size, which presents challenges for the sampling efficiency analysis. Denote the power of the binomial test~\eqref{eq:exact_bin_test} under the alternative where~$q_j>p_j$ by~$\cP(\psi_{j,n,\alpha})$. We have the following result pertaining relative sampling efficiency between tests conducted on two subpopulations:
\begin{theorem}\label{th:formal}
Consider the power of two exact binomial tests $\psi_{1,n_1,\alpha}$ and $\psi_{2,n_2,\alpha}$. Let $n_2 \in\bbN$, $p_2 \leq q_2 \leq 1/2$, $p_2 < p_1$, $\alpha \in (0,1)$, and $\cP(\psi_{2,n_2,\alpha})$ be such that
\begin{equation}\label{eq:cond-alpha}
\alpha \leq \tfrac{1}{2}-\sqrt{\frac{1}{2en_2p_2(1-p_2)}} \ , \qquad  \cP(\psi_{2,n_2,\alpha}) \geq \tfrac{1}{2} + \sqrt{\frac{2}{en_2q_2(1-q_2)}} \ .
\end{equation}
If
\begin{equation}\label{eq:gen-req-pq}
\frac{q_1}{q_2} \geq \frac{p_1}{p_2} \ ,
\end{equation}
then under the reduced sample size
\begin{equation}\label{eq:gen-req-n}
n_1 = \frac{q_2}{q_1}\cdot n_2 \ ,
\end{equation}
the powers of the two tests satisfy
\begin{equation}\label{eq:power-monotonic}
\cP(\psi_{1,\floor{n_1},\alpha}) > \cP(\psi_{2,{n_2},\alpha}) - \sqrt{\frac{q_1}{2e\floor{n_1}(1-q_1)}}\ind{ n_1 \text{ is not integer } } \ .
\end{equation}
\end{theorem}

The proof is deferred to Section~\ref{app:proofs}. Before we unpack and interpret the details of Theorem~\ref{th:formal}, we first highlight that a simple way to interpret the result is as a form of \emph{reassurance}. Specifically, although a normal approximation to the binomial test~\eqref{eq:exact_bin_test} may not necessarily accurately reflect its finite-sample power curves (see Figure~\ref{fig:power}), the sampling efficiency result one obtains under exact finite sample analysis \emdash i.e. the relation~\eqref{eq:gen-req-n} \emdash reassuringly \emph{coincides} with the sampling efficiency results one would obtain through such normality approximations (albeit under certain conditions, and with some degree of slack). What Theorem~\ref{th:formal} offers, then, are conditions~\eqref{eq:cond-alpha} such that normality-based approximations to sampling efficiency continue to approximately hold, along with a quantification of the slack in the difference between the two tests in~\eqref{eq:power-monotonic}.

We now interpret Theorem~\ref{th:formal} with respect to the surveillance context of this section. Theorem~\ref{th:formal} quantifies, under assumption~\eqref{eq:cond-alpha} and~\eqref{eq:gen-req-pq}, given a sample size $n_2$ for the test based on subpopulation 2, how small the sample size for the test based on subpopulation 1 can be chosen~\eqref{eq:gen-req-n}, for the two tests to retain nearly identical power. The required sample for subpopulation 1 is smaller than the fraction $p_2/p_1$ of $n_2$ since~\eqref{eq:gen-req-pq} and~\eqref{eq:gen-req-n} imply $n_1 \leq (p_1/p_2)n_2$. This ultimately implies that sampling from the subpopulation with high baseline prevalence is considerably more efficient if the baseline prevalence ratio~$p_2/p_1$ is small. 

Note that~\eqref{eq:gen-req-pq} holds if we assume that for both subpopulations the fraction of diseased subjects~$q_j$ under the alternative are a function of the baseline prevalence: 
\[
q_j = f(p_j) \ ,
\]
where $f$ is a nondecreasing superlinear function (i.e. $f(x)/x$ is nondecreasing in $x$). For example, if the outbreak prevalences are proportional to the baseline prevalence as $q_j = \nu p_j$ with $\nu > 1$ identical across the subpopulations.

\begin{rem}\label{rem:ump}
It may seem Theorem~\ref{th:formal} could be reinterpreted to concern the uniformly most powerful test, by applying the Karlin-Rubin Theorem \citep{Karlin1956}. This is incorrect, however, since the exact binomial test has exact type-I error only at a discrete set of significance levels (which depend on the null). At other significance levels the test is conservative. The uniformly most powerful test in this context is a randomized test, which are typically undesirable in practice and thus excluded from our discussion.
\end{rem}

\section{Monitoring cases in a sample}\label{sec:dyn}
We extend the static setting of the previous section to a monitoring setting to show how the previously obtained results may be applied and generalized. As before, our aim is to describe the relative sampling efficiency between samples from distinct subpopulations. Unlike the static setting where each subject $i$ in subpopulation $j$ has a fixed diseased status $x_{i,j}$, we now assume subjects have a independent recurrent (subject-specific) probability of getting diseased which we denote by $\omega_{i,j}$. For the purposes of exposition, we assume a ``daily'' probability of getting diseased, but any monitoring frequency can be substituted. In the context of this monitoring setting, subpopulations with a higher disease risk will have subjects with larger daily probabilities of getting diseased.

We observe the sequence of all daily newly diseased subjects within a sample of size $n_j$ from subpopulation~$j$, which we assume is sampled uniformly at random over the subpopulation prior to the monitoring period, and remains fixed throughout the monitoring period. Denote the set of sampled subject indexes by $I_j$ and the sequence of daily cases from this sample $I_j$ by $D_{j,1},D_{j,2},\dots | I_j$. If we assume no dependencies between the sampled subjects, and assume subjects may get diseased multiple times over the time horizon independently, then for all $t\in\mathbb{N}$ the daily observed cases can be modelled as independent sums of Bernoulli variables, i.e. $D_{j,t} | I_j  = \sum_{i \in I_j} B_{i,j,t}$ with $B_{i,j,t} \sim  \text{Bernoulli}(\omega_{i,j})$. In this setting, one wishes to use the sequence of observed cases $D_{j,1},D_{j,2},\dots | I_j$ to infer if the average rate of subjects getting diseased within subpopulation $j$, which we denote by $\omega_j = m_j^{-1}\sum_{i=1}^{m_j} \omega_{i,j}$, is larger than some pre-specified non-outbreak rate~$\lambda_{j}$. In other words, we wish to test if
\begin{equation}\label{hyp:dyn}
H_0^{(j)}: \omega_j = \lambda_j \ , \quad\text { vs. }\quad H_1^{(j)}: \omega_j > \lambda_j \ . 
\end{equation}
With knowledge of the sample size $n_j$, testing this hypothesis through a sequential application of the exact binomial test may be appropriate. This approach directly inherits the sampling efficiency given by Theorem~\ref{th:formal}, implying the same sampling efficiency conclusions in this monitoring setting.

However, in such settings one may not have knowledge of $n_j$ and $\lambda_j$ individually, but only of their product $n_j\lambda_j$ (i.e. the mean daily cases of the sample of size $n_j$ under non-outbreak circumstances). Methodology is then typically based on Poisson distributions; for example Poisson CUSUM control charts \citep{Brook1972, Rogerson2004, Jiang2013}. This approach can be motivated through the following approximations which we state here informally:
\begin{equation}\label{eq:poisson-approx}
D_{j,t} | I_j  = \textstyle\sum_{i \in I_j} B_{i,j,t} \overset{D}{\approx} P\Big( \textstyle\sum_{i \in I_j}\omega_{i,j} \Big) \overset{D}{\approx} P\left( n_j\omega_j \right)\ ,
\end{equation}
where $P(\mu)$ denotes a Poisson random variable with mean $\mu$ and $\overset{D}{\approx}$ (informally) denotes approximate equality of distribution. In~\eqref{eq:poisson-approx}, the first approximation follows from e.g. \citet[Theorem 1]{LeCam1960}; informally, this approximation (and the ensuing methodology) is then suitable when $n_j$ is large while the probabilities $\omega_{i,j}$ are small. In the second approximation, the random rate (induced by the sample~$I_j$) is replaced by its expectation. Informally, this approximation is suitable if the probabilities $\omega_{i,j}$ do not strongly vary between subjects. If these approximations are unsuitable, a potential alternative approach models the observations through a negative binomial distribution; this plays a role in Section~\ref{sec:case_study}.

Importantly, under the above approximations, the distribution of $D_{j,t} | I_j $ only depends on the product $n_j\lambda_j$ under the null. The CUSUM statistic can be defined recursively as
\[
C_t = \text{max}\left(0, C_{t-1} + D_{j,t} - k\right) \ ,
\]
where $k > 0$ is a reference value typically chosen such that $C_t$ is a scaled version of Wald's sequential probability ratio test \citep{Lorden1971, Lucas1985}. The null in~\eqref{hyp:dyn} is rejected (and an outbreak alarm raised) when $C_t$ exceeds a prescribed alarm threshold $h$. The run length of a control chart is defined as the time until the chart signals an alarm. One calibrates $h$ such that the average run length (ARL) is suitably large under the null, while being small under the alternative. 

We now obtain results analogous to Theorem~\ref{th:formal} for the Poisson CUSUM control chart. An important distinction between the analysis of Theorem~\ref{th:formal} and the arguments here is that (under the approximations in~\eqref{eq:poisson-approx}) one can directly relate the \emph{distributions} of the samples, rather than the behavior of the testing methodology. As a result, the discrete nature of the data plays a less prominent role. Nonetheless, some nuances remain pertaining the discreteness of the sample size $n_j$ \emdash these will surface in~\eqref{eq:alt-seq}.

\paragraph{Approximate result.} Because of the informal approximations made, we refrain from establishing a formal theorem and state the (approximate) results outright. As in the static setting, consider two subpopulations where the baseline risk is larger in subpopulation 1, i.e. $\lambda_1 > \lambda_2$ and consider sampling efficiency and outbreak rate assumptions akin to ~\eqref{eq:gen-req-pq} and~\eqref{eq:gen-req-n} as:
\begin{equation}\label{eq:sam-eff-dyn}
n_1 = \Big\lfloor \frac{\omega_2}{\omega_1} \cdot n_2 \Big\rfloor\ , \qquad \frac{\omega_1}{\omega_2} \geq \frac{\lambda_1}{\lambda_2} \ .
\end{equation}
Note the above implies $n_1 \leq \floor{(\lambda_2)/\lambda_1)n_2} \leq n_2$. With the above assumptions we now show (under approximations) that the reduced sample from subpopulation 1 leads to a shorter average detection time compared to the larger sample from subpopulation 2.

\begin{rem}
Akin to the results of the static setting, the assumption in the second equation of~\eqref{eq:sam-eff-dyn} holds if the probability of subjects getting diseased $\omega_j$ under the alternative are a nondecreasing function of the baseline probability $\lambda_j$: 
\begin{equation}\label{eq:outbreak-rate}
\omega_j = f(\lambda_j) \ , \text{ where $\frac{f(x)}{x}$ is nondecreasing in $x$} \ .
\end{equation}
\end{rem}

\paragraph{Justification of the approximate result.} For notational convenience, for two random variables $A$ and $B$, denote $A\preccurlyeq B$ if $B$ stochastically dominates $A$. Then, under the choice of~$n_1$ in~\eqref{eq:sam-eff-dyn} we have that, under the null hypothesis in~\eqref{hyp:dyn} and approximation~\eqref{eq:poisson-approx} for any $t\in\mathbb{N}$
\begin{equation}\label{eq:null-equivalence}
D_{1,t} | I_1 \overset{D}{\approx} P\left(n_1\lambda_1 \right) \preccurlyeq P\Big(\frac{\lambda_2}{\lambda_1} n_2 \lambda_1 \Big) \overset{D}{\approx}  D_{2,t} | I_2 \ .
\end{equation}
Thus under the reduced sample choice in~\eqref{eq:sam-eff-dyn}, observations under the null from subpopulation~1 are stochastically dominated by observations from subpopulation~2. This stochastic domination implies that alarm thresholds~$h$ which bound the ARL of the CUSUM statistic under the null for observations from subpopulation~2, also conservatively bound that same quantity for the CUSUM statistic from the reduced sample from subpopulation~1. This can be shown via a coupling argument \citep{Heidema2024}. Under the alternative hypothesis where $\omega_j > \lambda_j$ and approximation~\eqref{eq:poisson-approx} we get the converse of the above (approximate) stochastic domination:
\begin{equation}\label{eq:alt-seq}
D_{1,t} 
\overset{D}{\approx} P\left(n_1\omega_1 \right) 
= P\Big(\Big\lfloor \frac{\omega_2}{\omega_1}n_2\Big\rfloor\omega_1 \Big) 
\succcurlyeq P\Big(n_2\omega_2 - \omega_1 \Big)
\approx P\left(n_2\omega_2\right) 
\overset{D}{\approx} D_{2,t} \ ,
\end{equation}
where the second equality is due to~\eqref{eq:sam-eff-dyn} and the stochastic domination due to the bound $\floor{x} \geq x-1$. The final approximation in~\eqref{eq:alt-seq} is due to rounding in the definition of $n_1$ in~\eqref{eq:sam-eff-dyn} and holds exactly if the non-rounded quantity in~\eqref{eq:sam-eff-dyn} is an integer. The above implies under the alternative in~\eqref{hyp:dyn} \emdash i.e. under the outbreak scenario \emdash the observations from the small sample from subpopulation~1 (approximately) stochastically dominate the observations from the large sample from subpopulation~2. Since the alarm thresholds can be chosen identically by the arguments from~\eqref{eq:null-equivalence} this implies (if the approximation above would be exact) the ARL in an outbreak scenario is shorter or equal for the smaller sample from subpopulation~1. In other words, this implies a sampling efficiency result akin to Section~\ref{sec:settings-static}, that (if the approximations above would be exact) the smaller sample from subpopulation~1 leads to earlier or similar detection times of an outbreak compared to the larger sample from subpopulation~2. 

An exact characterization of how the approximations above \emdash particularly the final approximation in~\eqref{eq:alt-seq} \emdash propagate to the ARL is outside of the scope of this paper. Such an analysis is challenging due to the sequential dependencies of the CUSUM statistic. Related literature merely analyses the ARL numerically, typically through implicit formulations following Markov chain constructions \citep{Lucas1985, Hawkins1998}, and remains an interesting but challenging avenue for future work.

\section{Case study: COVID-19}\label{sec:case_study}
We highlight the usefulness of the above insights through a case study of daily COVID-19 infections in the Netherlands between June 1st, 2020 and July 31st, 2020. We observe the number of reported cases within several age cohorts~\citep{rivm}. At the start of this period, free PCR testing for anyone with mild symptoms was made available and the number of cases was low and stable \citep{rivm_updates}. It was relevant to monitor the reported case numbers for evidence of another infection wave potentially warranting interventions. Critically, in this stable period the number of cases was not zero: this fundamentally separates the setting of this case study from the setting of emerging diseases as considered by earlier works such as \cite{Ferguson2014, Mastin2017} discussed in Section~\ref{sec:intro}.

We consider the observation model described at the start of Section~\ref{sec:dyn} where we observe all daily reported cases within a sample from subpopulation $j$ to test the hypothesis~\eqref{hyp:dyn}. To show how the insights from preceding sections may continue to hold when more complex monitoring approaches are used, we slightly generalize the monitoring approach. Instead of modeling the daily number of reported cases $D_{j,t}$ through a Poisson distribution as in~\eqref{eq:poisson-approx}, we model these using a negative binomial distribution\footnote{For the uninitiated, the usage of the negative binomial distribution may seem somewhat arbitrary, but can be motivated from similar foundations as the Poisson distribution in~\eqref{eq:poisson-approx}. Specifically, with reference to the notation in Section~\ref{sec:dyn}, if the parameters $\omega_{i,j}$ are instead random and independently sampled from a Beta($a_1,a_2$) distribution then the resulting sum of $n$ terms has (under some conditions on the parameters $a_1$ and $a_2$) a limiting negative binomial distribution for $n\to\infty$. This specific result is derived in \cite{Boswell1970}. In \citet[Section 5.3]{Johnson1992}, the negative binomial as a limiting distribution is also discussed via other mixtures of distributions; e.g. as a mixture of the Poisson distributions, where their parameters follow a Gamma distribution.} with mean $n_j\omega_j$ and dispersion parameter $\theta_j$ which accounts for potential overdispersion of the observed counts (i.e. where the approximations in~\eqref{eq:poisson-approx} are inaccurate and the ensuing variance of $D_{j,t}$ is larger than its mean). Our goal remains testing the hypothesis~\eqref{hyp:dyn}, in this case corresponding to detecting a difference in the rate of reported cases. A suitable methodology is then the negative binomial Generalized Likelihood Ratio (GLR) control chart~\citep{Hohle2008}. 

We contrast the effectiveness of monitoring based samples from distinct subpopulations. As subpopulations, we consider five demographic age cohorts: age 20-29, age 30-39, age 40-49, age 50-59, and age 60-69. In the remainder these are indexed by $j=1,\dots,5$ respectively.
We consider the first three weeks of the data (June 1st -- June 21st) as a baseline period. The observed average daily reported cases within each subpopulation in this period are used to estimate the null mean case numbers $n_j\lambda_j$ and dispersion parameters $\theta_j$ for $j=1,\dots,5$. The values are given in Table~\ref{tb:baseline_pars} and we assume these values as fixed for monitoring starting from the 22nd of June. Incidentally, the average is highest for the youngest subpopulation and decreases for the older subpopulations.

\begin{table}[ht]
\centering
\begin{tabular}{@{}l>{\raggedright\arraybackslash}p{3cm}>{\raggedright\arraybackslash}p{3cm}>{\raggedright\arraybackslash}p{4cm}@{}}
\toprule
Subpopulation & Estimated mean cases under the null $n_j\lambda_j$ & Estimated overdispersion $\theta_j$  & Demographic size in millions in 2020 \citep{cbs} \\ \midrule
Age 20-29 & 22.9 & 0.075 & 2.23\\
Age 30-39 & 21.6 & 0.016 & 2.15\\
Age 40-49 & 19.5 & 0.058 & 2.21\\
Age 50-59 & 19.5 & 0.162 & 2.53\\
Age 60-69 & 10.3 & 0.116 & 2.11\\ \bottomrule
\end{tabular}
\caption{Estimated non-outbreak parameter values for the model of the daily number of observed cases $D_{j,t}$ for the subpopulations considered in Section~\ref{sec:case_study}. In the non-outbreak scenario (between June 1st - June 21st) the daily number of observed cases are modeled as independent and identically distributed samples from a negative binomial distribution with mean $n_j\lambda_j$ and variance $n_j\lambda_j + \theta_j(n_j\lambda_j)^2$. Estimation is based on maximum likelihood and implemented through the MASS package \citep{mass}. Mean estimates are rounded to one decimal, overdispersion estimates to three decimals, and the population size to two decimals.}\label{tb:baseline_pars}
\end{table}

Within this case study we do not control the sample size. Instead, we assume that the samples sizes from each of the subpopulations $j=1,\dots,5$ are of the same (but unknown) size (and these remain fixed throughout the monitoring period). This is a strong assumption which we motivate as follows: first, the population is evenly distributed across the considered age groups, with each group containing approximately the same number\footnote{As shown in Table~\ref{tb:baseline_pars} the age group 50-59 is actually slightly larger than the other groups; however, if this would imply that the sample underlying age 50-59 is larger, then this would strengthen the observations in this case study; we discuss this at the end of Section~\ref{sec:case_study}.} of individuals \citep{cbs} \emdash included in Table~\ref{tb:baseline_pars}. Secondly, apart from individuals gaining knowledge about being potentially infected with COVID-19, there was no specific additional incentive for individuals to get tested at any point during this period; for example, there was no policy in place that restricted individuals from attending events unless they could show a recent negative test. The interpretation of this case study critically depends on this assumption (as well as others). We reflect on this assumption at the end of this section. 

With the observed averages in Table~\ref{tb:baseline_pars} the above assumption implies a decreasing ordering on the (estimated) non-outbreak average disease rates, where $\lambda_1 > \lambda_2 > \lambda_3 \approx \lambda_4 > \lambda_5$. Now, we assume that the outbreak rates satisfy~\eqref{eq:outbreak-rate}. Then by the arguments of Section~\ref{sec:dyn} we expect (if~\eqref{eq:poisson-approx} is accurate and a Poisson CUSUM chart is used) on average, monitoring based on the younger subpopulations to detect outbreaks earlier than based on older subpopulations, since the underlying samples are of the same size but the younger subpopulations have a higher sampling efficiency. Here, we investigate if this result is observable in the data arising in this case study where the negative binomial GLR control chart is used.

\begin{figure}[htbp]
\centering
\begin{subfigure}{0.7\textwidth}
\includegraphics[width=\textwidth]{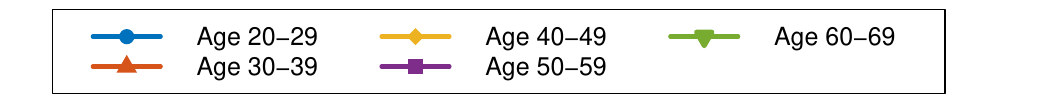}\vspace{0.3cm}
\end{subfigure}

\begin{subfigure}{0.9\textwidth}
\includegraphics[width=\textwidth]{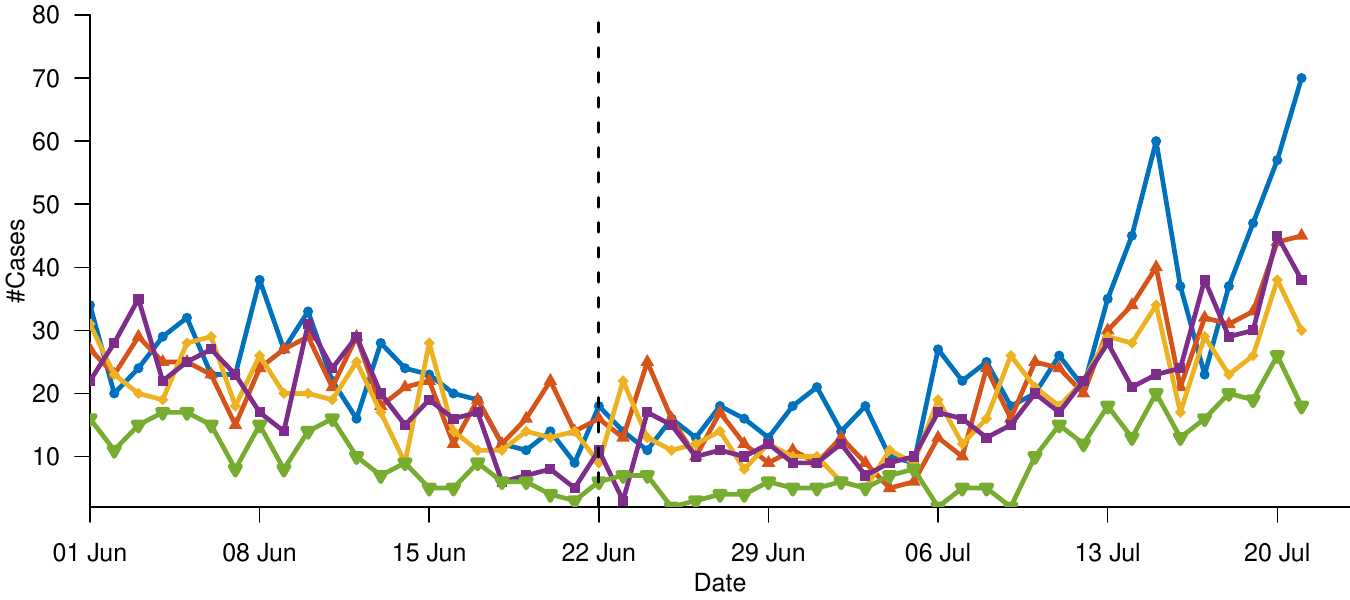}\caption{\justifying Daily COVID-19 case numbers \citep{rivm}, stratified by age cohort. The vertical line at 22nd of June marks the start of the monitoring period.} 
\end{subfigure}
\begin{subfigure}{0.9\textwidth}
\includegraphics[width=\textwidth]{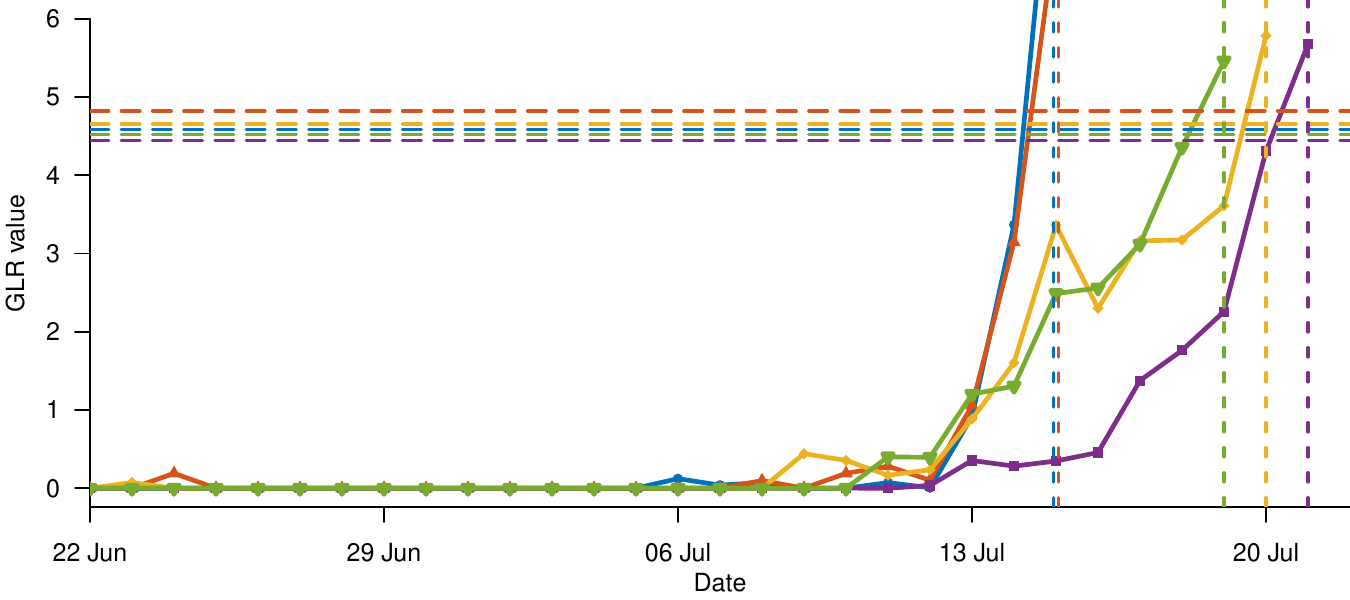}\caption{\justifying GLR statistic at each time point starting from 22nd of June. Horizontal lines indicate alarm thresholds of the charts. Vertical lines indicate the time of first alarm. The GLR statistic is computed through the surveillance package \citep{Salmon2016}. Results presented use a window-limited GLR control chart with a width of 20 datapoints, and set up to detect increases in the mean of the observations with respect to the null parameters estimated as in Table~\ref{tb:baseline_pars}. Alarm thresholds were computed through Monte Carlo approximation with $10^4$ repetitions such that the average run length under the null is at least 370 days. }
\end{subfigure}
\caption{Graphical results discussed in Section~\ref{sec:case_study}.}\label{fig:C_plots}
\end{figure}

Case numbers and the GLR monitoring statistic are plotted in Figure~\ref{fig:C_plots}. Numerical studies are conducted using R \citep{RCoreTeam2024} and the surveillance package \citep{Salmon2016}. The results corroborate the insights from previous sections. The monitoring statistics on the younger subpopulations (with high baseline risk) raise alarms earlier than those based on the older subpopulations, nearly ordered by the ordering of their baseline rate $\lambda_j$ \emdash i.e. the younger subpopulations indeed correspond to the proverbial ``canaries'' within this case study. One exception to the results we expect is the subpopulation of age 60-69, which raises an alarm earlier than the age groups 40-49 and 50-59, while our insights suggest it should raise an alarm later than all the other subpopulations. This could be due to the lack of validity of model assumptions such as~\eqref{eq:outbreak-rate}. However, one should note our theoretical results pertain only \emph{average} behavior, so it is not unexpected that for some sample realizations the alarms are not ordered by their average ordering. We further illustrate this via a simulation study in Appendix~\ref{app:simstudy}. In Figure~\ref{fig:C_plots} thresholds are chosen such that the ARL under the null is at least 370 days, matching the Shewart control chart with default settings \citep{Hawkins1998}. For other threshold values, the time of the alarms naturally shift but the insights do not qualitatively change.

\paragraph{Discussion.} We have made various simplifying assumptions within this case study, and naturally our conclusions depend on their validity. One critical assumption in this case study was the assumption of equal sample sizes underlying the age groups. We conclude by discussing the interpretation of the above results if one is not willing to make this assumption. 

First, recall from the model of Section~\ref{sec:dyn} that the sample size within our monitoring context refers to all individuals in the monitored population which will report their diseased status. The underlying sample size is therefore dependent on the fraction of individuals who, should they get infected, will get themselves tested. For the individual, this may depend various factors including how severely the infection manifests. A more complex model could approximate this through additional randomness. However, the parameters within such a process are not straightforward to estimate and a more in-depth investigation is outside of the scope of this paper. 

If one is unwilling to assume the underlying sample size are of the same size, and instead believes that older subpopulations have a larger sample underlying their observed cases, then the conclusions of this case study are stronger: this implies a \emph{smaller} sample of the younger subpopulation leads to earlier detection. Conversely, if one believes that the observed cases of the younger age groups arise from a larger underlying sample than those of the older age groups, then the implications of the case study are unclear. In that case, the observed ``canary effect'' of the younger age groups may then instead be explained partially or completely by the increased sample size. 

\section{Final remarks}\label{can:sec:discussion}
Practitioners should be mindful that (under the assumptions we have considered) the most efficient subpopulations are those where individuals have the highest probability of contracting the disease under study; this may not necessarily coincide with subpopulations most at risk of \emph{developing complications}. Identification of subpopulations could be based on domain knowledge (e.g. through subgroups of suitable subject covariates) or, as exemplified in Section~\ref{sec:case_study}, based on historical data if accurate risk estimates can be obtained.

We have shown how these efficiency results can be observed in practice through a case study in Section~\ref{sec:case_study}. Future work could discuss other case studies; for example the surveillance of international travellers as discussed in Section~\ref{sec:intro}. Furthermore, in practice one may not be fully in control of the sampling strategy, and one may be faced with samples from suboptimal subpopulations. Finding practicable methodology to optimally combine such suboptimal samples, particularly in the presence of overdispersion of observed case numbers, is another interesting avenue for future work.

\FloatBarrier
\section{Mathematical derivations}\label{app:proofs}
\subsection{Proof of Theorem~\ref{th:formal}}\label{sec:proof-th-formal}
\begin{proof}
Recall the exact binomial test under study as in~\eqref{eq:exact_bin_test}. We aim to lower bound the power of this test, when based on samples from subpopulation 1, by the power of the test under samples from subpopulation 2. Its power can be expressed as
\[
\cP\left(\psi_{1,\floor{n_1},\alpha}\right) = 1-F_{\floor{n_1},q_1}\left(F^{-1}_{\floor{n_1},p_1}(1-\alpha) \right) \ ,
\]
such that a lower bound can be obtained by finding an upper bound for both $F_{\floor{n_1},q_1}(x)$ and $F^{-1}_{\floor{n_1},p_1}(1-\alpha)$. The proof hinges on bounds and monotonicity results of the binomial cumulative distribution function, where one must carefully account for the discreteness of the distributions involved. Before proceeding with the proof, we first state the technical results needed. A standard result is that
\begin{equation}\label{eq:bin-dom-prob}
F_{n,u}(x) \leq F_{n,v}(x) \ ,
\end{equation}
for all $x\in\bbR$ when $u \geq v$, which can be shown through a stochastic domination argument. A crucial and more technical result is the following lemma, which is a minor adaptation from the results derived in~\cite{Anderson1967}:
\begin{lemma}[adapted from~\cite{Anderson1967}]\label{lem:monotonicity}
Let $n/a \in \bbN$ and $a \geq 1$. Then:
\begin{align*}
F_{n,p}(x) &< F_{\frac{n}{a},ap}(x) \quad \forall x \geq np  \ ,\\
F_{n,p}(x) &> F_{\frac{n}{a},ap}(x) \quad \forall x \leq np-1 \ .
\end{align*}
\end{lemma}
The proof is provided at the end of this section.  We also require a number of other technical results, which we collect in the following lemma and prove at the end of this section:

\begin{lemma}\label{lem:binom-tech}
\hspace{0.01cm}\begin{enumerate}[label=(\roman*)]
\item For all $x\in\bbR$: 
\[
F_{n,p}(x) - F_{(n+1),p}(x) \leq \sqrt{\frac{p}{2en(1-p)}} \ .
\] \label{en:lem-binom-tech-2cdf}
\item If $n/a \in \bbN$ and $a \geq 1$, then:
\[
F_{n,p}^{-1}(u) \geq F_{\frac{n}{a},ap}^{-1}(u) \quad \forall u \geq \tfrac{1}{2} + \sqrt{\tfrac{1}{2enp(1-p)}} \ .
\] \label{en:lem-binom-tech-quantile}
\item For all $n$ and $p$ we have
\begin{align*}
F^{-1}_{n,p}\left(\tfrac{1}{2}-\sqrt{\tfrac{2}{enp(1-p)}}\right) &\leq np - 1 \ ,\\
F^{-1}_{n,p}\left(\tfrac{1}{2}+\sqrt{\tfrac{1}{2enp(1-p)}}\right) &\geq np \ .
\end{align*} \label{en:lem-binom-tech-quasimedian} 
\end{enumerate}
\end{lemma}

We start by bounding $F_{\floor{n_1},q_1}(x)$. First note if $x \leq n_2q_2-1$ we have
\begin{equation}\label{eq:dom-cdf}
F_{\ceil{n_1},q_1}(x) = F_{\frac{n_2}{n_2/\ceil{n_1}},q_1}(x) \leq F_{\frac{n_2}{n_2/\ceil{n_1}},\frac{q_2n_2}{\ceil{n_1}}}(x) < F_{n_2,q_2}(x) \ ,
\end{equation}
where the first inequality follows from~\eqref{eq:bin-dom-prob} since assumptions~\eqref{eq:gen-req-n} and~\eqref{eq:gen-req-pq} ensure 
\[
\ceil{n_1} \geq n_1 = \frac{q_2n_2}{q_1} \Longleftrightarrow q_1 \geq \frac{q_2n_2}{\ceil{n_1}} \ ,
\]
and the second inequality follows from application of the second result of Lemma~\ref{lem:monotonicity} with $a = n_2/\ceil{n_1} \geq 1$. We can then bound our quantity of interest by observing that for $x \leq n_2q_2 - 1$:
\begin{align*}
F_{\floor{n_1},q_1}(x) &= F_{\ceil{n_1},q_1}(x) + \Bigg(F_{\floor{n_1},q_1}(x)  - F_{\floor{n_1}+1,q_1}(x)\Bigg)\ind{ \floor{n_1} \neq \ceil{n_1}} \\ 
&< F_{n_2,q_2}(x) + \sqrt{\frac{q_1}{2e\floor{n_1}(1-q_1)}}\ind{ \floor{n_1} \neq \ceil{n_1}} \ , \numberthis\label{eq:bound-cdf-floor}
\end{align*}
where we used~\eqref{eq:dom-cdf} and result~\ref{en:lem-binom-tech-2cdf} from Lemma~\ref{lem:binom-tech}. Next, we bound the quantile $F_{\floor{n_1},p_1}^{-1}(1-\alpha)$ as
\begin{equation}\label{eq:crit-value}
F_{\floor{n_1},p_1}^{-1}(1-\alpha) = F_{\frac{n_2}{n_2/\floor{n_1}},p_1}^{-1}(1-\alpha) \leq F_{\frac{n_2}{n_2/\floor{n_1}},\frac{p_2n_2}{\floor{n_1}}}^{-1}(1-\alpha) \leq F_{n_2,p_2}^{-1}(1-\alpha) \ ,
\end{equation}
where the first inequality follows from~\eqref{eq:bin-dom-prob} since assumption~\eqref{eq:gen-req-n} ensures
\[
\floor{n_1} \leq n_1 = \frac{p_2n_2}{p_1} \Longleftrightarrow p_1 \leq \frac{p_2n_2}{\floor{n_1}} \ ,
\]
and the second inequality follows from result~\ref{en:lem-binom-tech-quantile} from Lemma~\ref{lem:binom-tech} with $a = n_2/\floor{n_1} \geq 1$ since~\eqref{eq:cond-alpha} implies $1-\alpha \geq \tfrac{1}{2}+1/\sqrt{2en_2p_2(1-p_2)}$. As our bound~\eqref{eq:bound-cdf-floor} is only valid for $x \leq n_2q_2-1$ we must ensure the bound on our quantile~\eqref{eq:crit-value} is small enough such that we can use both bounds in unison. Since the power of the test $\cP(\psi_{2,n_2,\alpha}) \geq \frac{1}{2}+\sqrt{2/(en_2q_2(1-q_2))}$ by~\eqref{eq:cond-alpha} we can upper bound the quantile $F_{n_2,p_2}^{-1}(1-\alpha)$ by
\begin{align*}
\quad &\cP(\psi_{2,n_2,\alpha}) \geq \frac{1}{2}+\sqrt{\frac{2}{en_2q_2(1-q_2)}} \\
\Longrightarrow\quad& F_{n_2,q_2} \Big( F_{n_2,p_2}^{-1}(1-\alpha)\Big) \leq \frac{1}{2}-\sqrt{\frac{2}{en_2q_2(1-q_2)}} \\
\Longrightarrow\quad& F_{n_2,p_2}^{-1}(1-\alpha) \leq F_{n_2,q_2}^{-1}\left(\frac{1}{2} - \sqrt{\frac{2}{en_2q_2(1-q_2)}}\right) \leq n_2q_2 - 1 \ , \numberthis\label{eq:crit-bound}
\end{align*}
where in the last inequality we used~\ref{en:lem-binom-tech-quasimedian} from Lemma~\ref{lem:binom-tech}. Having ensured~\eqref{eq:crit-bound}, we ultimately find
\begin{align*}
\cP(\psi_{1,\floor{n_1},\alpha}) &= 1-F_{\floor{n_1}, q_1}\Big( F_{\floor{n_1},p_1}^{-1}(1-\alpha) \Big) \\
&\geq 1-F_{\floor{n_1}, q_1}\Big( F_{n_2,p_2}^{-1}(1-\alpha) \Big) \\
&> 1-F_{n_2, q_2}\Big( F_{n_2,p_2}^{-1}(1-\alpha) \Big) -\sqrt{\frac{q_1}{2e\floor{n_1}(1-q_1)}}\ind{ \floor{n_1} \neq \ceil{n_1}} \\
&= \cP(\psi_{2,n_2,\alpha}) - \sqrt{\frac{q_1}{2e\floor{n_1}(1-q_1)}}\ind{ \floor{n_1} \neq \ceil{n_1}} \ ,
\end{align*}
where the first inequality follows from~\eqref{eq:crit-value} and the second inequality follows from~\eqref{eq:bound-cdf-floor} which holds due to the result in~\eqref{eq:crit-bound}. 
\end{proof}

\subsection{Proof of Lemma~\ref{lem:monotonicity}}\label{app:monotonicity}
We adapt Theorem 2.1 from~\cite{Anderson1967}. Define $Y_i \sim \text{Bernoulli}(p_i)$ for $i=1,\dots,n$. Denote the CDF of their sum by $H(k) \equiv \P{ \sum_{i=1}^n Y_i \leq k}$ and $\lambda = \sum_{i=1}^n p_i$. Now~\citet[Theorem 4]{Hoeffding1956} implies
\begin{align*}
F_{n,\lambda/n}(k) &\geq H(k) \quad \forall k\leq \lambda-1 \ , \\
F_{n,\lambda/n}(k) &\leq H(k) \quad \forall k\geq \lambda \ .
\end{align*}
Particularize $\lambda = np$ with $p_1,\dots,p_{n - n/a} = 0$ and
\[
p_{n-n/a+1},\dots,p_n = \frac{np}{n/a} = ap \ .
\]
Under this particularization, $H(k) = F_{n/a, ap}(k)$, such that the results above imply the statement of the lemma.

\subsection{Proof of Lemma~\ref{lem:binom-tech}}\label{sec:lem:binom-tech}
We start by proving statement~\ref{en:lem-binom-tech-2cdf}. Denote $Y_i \sim \text{Bernoulli}(p)$, we have that for any $x\in\mathbb{R}$:
\begin{align*}
&F_{(n+1),p}(x) = \P{\sum_{i=1}^{n+1} Y_i \leq x} \\
&= \P{\sum_{i=1}^{n} Y_i \leq x-1}\P{Y_{n+1} = 1} + \P{\sum_{i=1}^{n} Y_i \leq x}\P{Y_{n+1} = 0} \\
&= pF_{n,p}(x-1) + (1-p)F_{n,p}(x) \\
&= p\Big(F_{n,p}(x) - f_{n,p}(\floor{x})\Big) + (1-p)F_{n,p}(x) \\
&= -pf_{n,p}(\floor{x}) + F_{n,p}(x) \ .
\end{align*}
Rearranging the quantity above and using the following result from \cite{Herzog1947}:
\begin{equation}\label{eq:upperbound-binomcdf}
f_{n,p}(x) \leq \sqrt{\frac{1}{2enp(1-p)}} \ ,
\end{equation}
completes the proof. Next we prove statement~\ref{en:lem-binom-tech-quasimedian}. We have that
\begin{align*}
F_{n,p}(np-1) &= F_{n,p}(\ceil{np}) - \Big(F_{n,p}(\ceil{np}) - F_{n,p}(\floor{np-1})\Big) \\
&\geq \tfrac{1}{2} - f_{n,p}(\ceil{np}) - f_{n,p}(\floor{np}) \\
&\geq \tfrac{1}{2} - 2\sqrt{\frac{1}{2enp(1-p)}} \ ,
\end{align*} 
where in the first inequality we used that the median of the binomial distribution is smaller than $\ceil{np}$, and in the second inequality we used~\eqref{eq:upperbound-binomcdf}. Rearranging the inequalities then implies the first result of statement~\ref{en:lem-binom-tech-quasimedian}. Now, denote $Z_{n,p}$ a binomial random variable with count $n$ and probability $p$, such that analogously we obtain
\begin{align*}
1-F_{n,p}(np) &= \P{Z_{n,p} > np} \\
&= \P{Z_{n,p} \geq \floor{np}+1} \\
&= \P{Z_{n,p} \geq \floor{np}} - f_{n,p}(\floor{np}) \\
&\geq \tfrac{1}{2} - \sqrt{\frac{1}{2enp(1-p)}} \ ,
\end{align*} 
where we used that the median of the binomial distribution is larger than $\floor{np}$ and~\eqref{eq:upperbound-binomcdf}. Rearranging the inequality implies the second result of statement~\ref{en:lem-binom-tech-quasimedian}. Finally, we prove statement~\ref{en:lem-binom-tech-quantile}. Recall
\[
F^{-1}_{n,p}(u) = \inf\{ x\in\bbR : F_{n,p}(x) \geq u \} \ .
\]
Lemma~\ref{lem:monotonicity} implies that
\begin{align*}
&\forall x \geq np: \quad F_{n,p}(x) < F_{n/a,ap}(x) \\
\Longrightarrow \quad& \forall x \geq np \quad \forall u: \quad F_{n,p}(x) \geq u \Rightarrow F_{n/a,ap}(x) \geq u \\
\Longrightarrow \quad& \{x \geq np : F_{n,p}(x) \geq u \} \subseteq \{x \geq np : F_{n/a,ap}(x) \geq u \} \ .
\end{align*}
Now assume that $u \geq 1/2 + \sqrt{1/2enp(1-p)}$. Then setting  $F_{n,p}(x) \geq u$ and solving for $x$ implies (through monotonicity of $F_{n,p}$ and statement~\ref{en:lem-binom-tech-quasimedian} that:
\[
F_{n,p}(x) \geq u \Rightarrow x \geq F^{-1}_{n,p}(u) \geq np \ .
\]
Therefore under the assumption $u \geq 1/2 + \sqrt{1/2enp(1-p)}$ we have:
\[
\{x \geq np : F_{n,p}(x) \geq u \} = \{ x \in \bbR : F_{n,p}(x) \geq u \} \ ,
\]
and likewise 
\[
\{x \geq np : F_{n/a,ap}(x) \geq u \} = \{ x \in \bbR : F_{n/a,ap}(x) \geq u\} \ ,
\]
Finally this implies, if $u \geq 1/2 + \sqrt{1/2enp(1-p)}$ that
\begin{align*}
F_{n,p}^{-1}(u) &= \inf\{ x \in \bbR : F_{n,p}(x) \geq u \} \\
&\geq \inf\{x \in \bbR : F_{n/a,ap}(x) \geq u \} = F_{n/a,ap}^{-1}(u) \ ,
\end{align*}
which completes the proof.

\appendix
\section{Illustration of sampling variability of alarm ordering}\label{app:simstudy}
In the case study of Section~\ref{sec:case_study} we expected that the older subpopulation would lead to a longer average detection time (under the assumption that the results of Section~\ref{sec:dyn} would generalize to the more complex monitoring setting considered) compared to the younger subpopulations. However, we observe monitoring based on the subpopulation of age 60-69 raises an alarm earlier than based on the subpopulation age 40-49. Here, we illustrate that if the daily cases follow the negative binomial model as in Section~\ref{sec:case_study} with parameters as estimated in Table~\ref{tb:baseline_pars}, then indeed we expect \emph{on average} earlier detection times for the younger subpopulation, but the sampling variability will sometimes result in the detection time of the younger population to be longer.

To illustrate this, we simulate daily cases from subpopulation age 40-49 and age 60-69 (indexed with $j=3$ and $j=5$ as in Section~\ref{sec:case_study}). The model used to sample the observations corresponds to the assumed model in Section~\ref{sec:case_study} as:
\begin{align*}
D_{3,t} &\sim \text{NegativeBinomial}(\nu n_3\lambda_3, \theta_3) \ , \numberthis\label{eq:simulation}\\
D_{5,t} &\sim \text{NegativeBinomial}(\nu n_5\lambda_5, \theta_5) \ ,
\end{align*}
independently, where the parameters $n_j\lambda_j$ and $\theta_j$ are used as estimated in Table~\ref{tb:baseline_pars} and $\nu \geq 1$ is a free parameter. If $\nu = 1$ the above model corresponds to the assumed null model of the observations, and if $\nu > 1$ we are simulating under the alternative. We use the GLR control chart as in Section~\ref{sec:case_study} to monitor sequences of simulated observations from~\eqref{eq:simulation}.

\begin{figure}[htbp]
\centering
\includegraphics[width=0.6\textwidth]{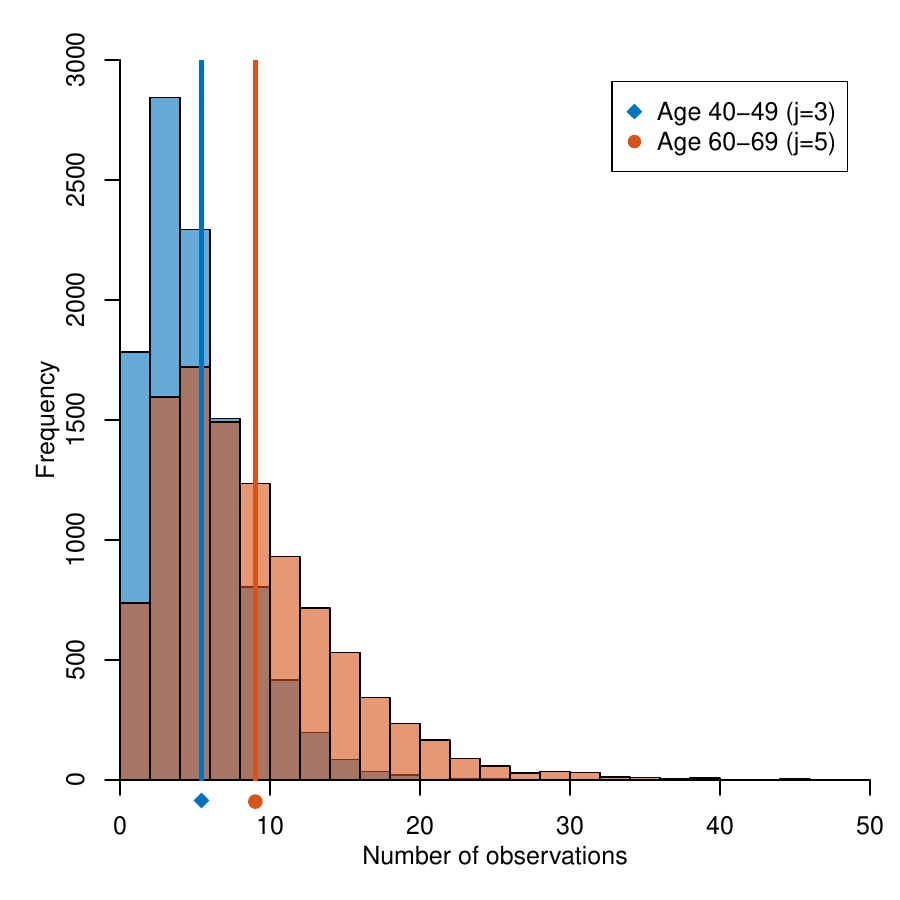}\caption{Histograms of the time until detection when a negative binomial GLR control chart is used on independent samples from~\eqref{eq:simulation} for $\nu=1.5$. Vertical lines indicate the observed average time until detection. The GLR statistic is computed through the surveillance package \citep{Salmon2016}. Results presented use a window-limited GLR control chart with a width of 20 datapoints, and set up to detect increases in the mean of the observations. The chart assumes the baseline behavior corresponds to the observation model~\eqref{eq:simulation} with $\nu=1$. where the parameters $n_j\lambda_j$ and $\theta_j$ are used as estimated as in Table~\ref{tb:baseline_pars}. Alarm thresholds were computed through Monte Carlo approximation with $10^4$ repetitions such that the average in-control run length is 370 days. The results are based on $10^4$ independent repetitions.}\label{fig:time_until_detection}
\end{figure}

For $\nu = 1.5$, histograms of the time until detection are plotted in Figure~\ref{fig:time_until_detection}. We observe two phenomena: first, as expected, the average time until first detection is indeed smaller for the younger subpopulation. Secondly, we can compute (within the context of this simulation) the probability of observing samples from~\eqref{eq:simulation} which lead to an earlier alarm from the older subpopulation compared to the younger subpopulation. This probability is estimated as $26.9\%$ (rounded to one decimal). Therefore, while (in this simulation setting) we indeed observe the younger subpopulation leads to earlier detection on average, it is not unexpected that one occasionally observes samples such as those in our case study.

\newpage

\phantomsection
\addcontentsline{toc}{section}{Acknowledgements}
\section*{Acknowledgements}
I would like to thank R.C. G\"undlach, D.H. Hamer, S. Heidema, E.R. van den Heuvel, R. Huits, K. O’Laughlin, M. Libman, R.A.J. Post, and M. Regis for useful discussions regarding earlier drafts of this manuscript. This project was funded through a cooperative agreement between the US Centers for Disease Control and Prevention and the International Society of Travel Medicine (Federal Award Number: 1U01CK000632-01-00).

\FloatBarrier
\phantomsection\addcontentsline{toc}{section}{References}
\small
\urlstyle{rm}
\interlinepenalty=10000
\bibliographystyle{custom_bib}
\bibliography{bib, extra}

\end{document}